\documentclass{PoS}

\usepackage{url}
\usepackage{graphicx}

\title{Synergies between astroparticle, particle and nuclear physics}

\ShortTitle{Synergies between astroparticle, particle and nuclear physics}

\author{\speaker{Caterina Doglioni}\thanks{Versions of this article have appeared on the CERN EP newsletter (Summer 2019) and on the ECFA newsletter (September 2019).}\\
        Lund University\\
        E-mail: \email{caterina.doglioni@hep.lu.se}}

\abstract{
One overarching objective of science is to further our understanding of the universe, from its early stages to its current state and future evolution. This depends on gaining insight on the universe's most macroscopic components, for example galaxies and stars, as well as describing its smallest components, namely elementary particles and nuclei and their interactions. It is clear that this endeavour requires combined expertise from the fields of astroparticle physics, particle physics and nuclear physics. 
Pursuing common scientific drivers also require mastering challenges related to instrumentation (e.g. beams and detectors), data acquisition, selection and analysis, and making data and results available to the broader science communities. Joint work and recognition of these foundational topics will help all communities grow towards their individual and common scientific goals. The \href{https://indico.cern.ch/event/577856/contributions/3467632/attachments/1880133/3097510/20180513_Doglioni_Synergies_EPS_small.pdf}{talk} corresponding to this contribution has been presented during the special ECFA session of EPS-HEP 2019 focused on the update of the European Strategy of Particle Physics.}

\FullConference{
European Physical Society Conference on High Energy Physics - EPS-HEP2019 -\\
			10-17 July, 2019\\
			Ghent, Belgium}

\begin{document}

\section{Introduction}

One overarching objective of science is to further our understanding of the universe, from its early stages to its current state and future evolution. This depends on gaining insight on the universe's most macroscopic components, for example galaxies and stars, as well as describing its smallest components, namely elementary particles and nuclei and their interactions. It is clear that this endeavor requires combined expertise from the fields of astroparticle physics, particle physics and nuclear physics.

A number of the contributions and discussions at the recent Granada meeting for the update of the European Strategy of Particle Physics, as well as the contribution at the EPS-HEP ECFA Open Session summarized in this newsletter, highlighted a growing wish for closer collaboration between ECFA and the astrophysics (APPEC, \url{https://www.appec.org}) and nuclear physics (NuPECC, \url{http://www.nupecc.org}) communities.

Many physics problems where synergies between particle physics, astrophysics and nuclear physics are required are discussed in the APPEC and NuPECC strategy documents (see links at the bottom of this piece). Among those, this contribution focused on the challenge of of elucidating the nature of 27\% of the matter-energy content of the universe, commonly called \textit{dark matter}. Pursuing these scientific goals also requires mastering challenges related to instrumentation (e.g. beams and detectors), data acquisition, selection and analysis, and making data and results available to the broader science communities. Joint work and recognition of these \textit{foundational} topics, also covered in detail in the contributions by C. Biscari, A. Cattai and G. A. Stewart in the ECFA newsletter~\cite{ECFANewsletter}, will help all communities grow towards their individual and common scientific goals. This contribution presented one of the many common challenges faced by particle physics and astrophysics: the necessity of dealing with large, sometimes heterogeneous datasets and derive insight from them in short periods of time.

\section{New physics discoveries and dark matter}

The Large Hadron Collider has yielded the discovery of a new particle, the Higgs boson. Precision measurements and fits of other quantities in the Standard Model of Particle Physics guided a search lasting decades after the conception of the Higgs mechanism. With the European Strategy Update, the particle physics community is currently deciding what are the best tools to employ to test the Standard Model: \textit{how/where to look next for new physics} is a relevant question in this process, given that hints coming from the Standard Model itself are not as telling as in the case of the Higgs boson. For this reason, research directions for physics beyond the Standard Model can be found in open problems in astrophysics that need a systematic exploration, for example the determination of the nature of dark matter.

One of the many explanations for this dark matter is that it is composed by new massive particles that interact only weakly with ordinary matter particles, or Weakly Interacting Massive Particles (WIMPs). These new particles can be produced at colliders, as well as detected by direct and indirect detection astrophysics experiments in space and underground (see Ref.~\cite{DarkMatterFeature} and links for a basic summary). By producing new particles in the lab, colliders are well placed to understand the nature of these particle’s interactions with ordinary matter. The necessary confirmation that these new particles have also a cosmological origin comes from complementary observations in direct and indirect detection experiments.

\begin{figure}[!htb]
        \center{\includegraphics[width=\textwidth]
        {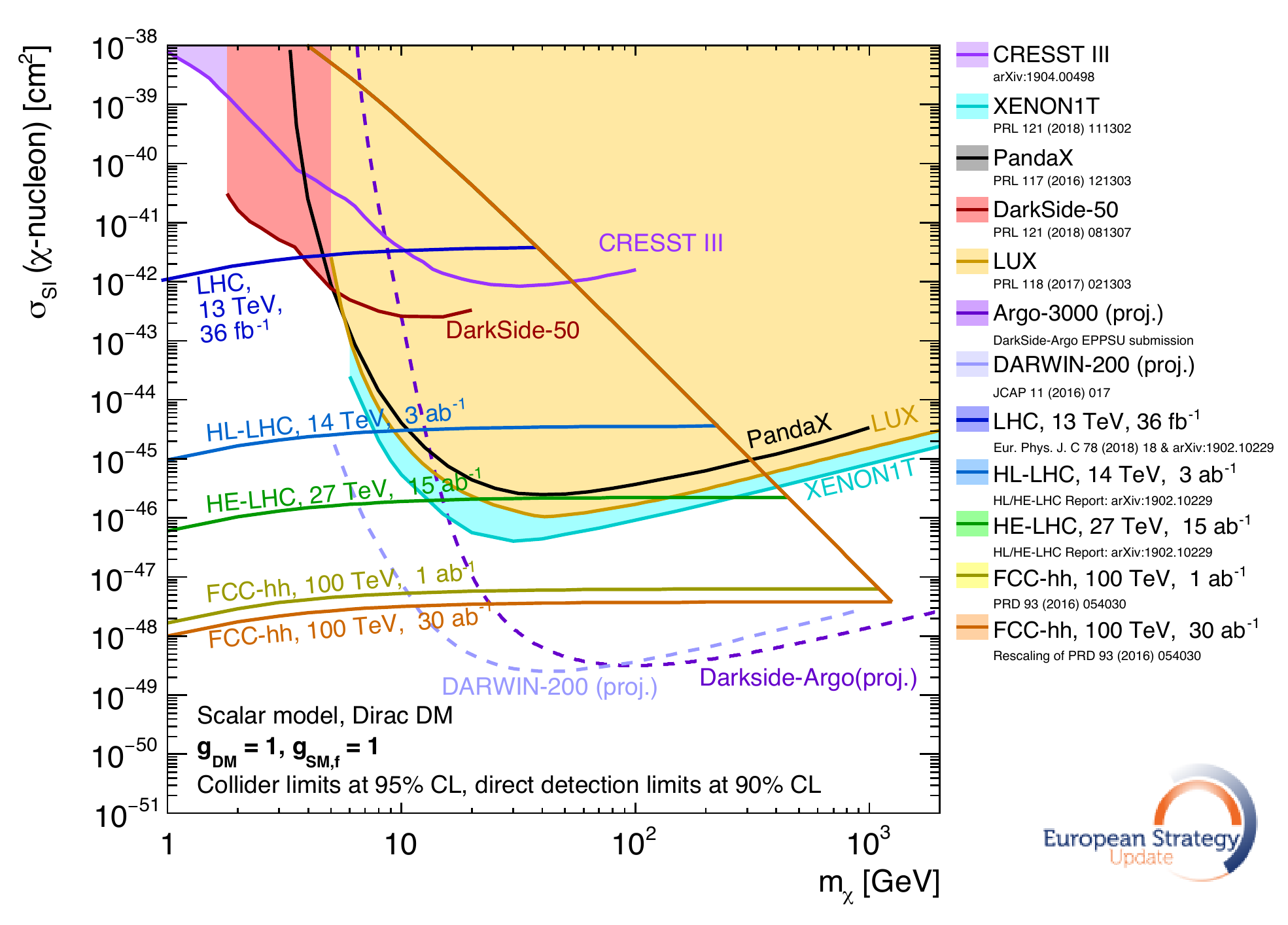}}
        \caption{\label{fig:DMCombinationDD} Comparison of sensitivities of future collider and direct detection experiments within a simplified model scenario of a WIMP where the interaction between Standard Model quarks and the dark matter is mediated by a new scalar particle. If dark matter is composed by particles within this model with a mass between 10 GeV and 1 TeV, future colliders and direct detection experiment can confirm each other’s discoveries in the next decades. Adapted from Ref.~\cite{Strategy:2019vxc}, with the addition of ATLAS results line.}
\end{figure}

\begin{figure}[!htb]
        \center{\includegraphics[width=\textwidth]
        {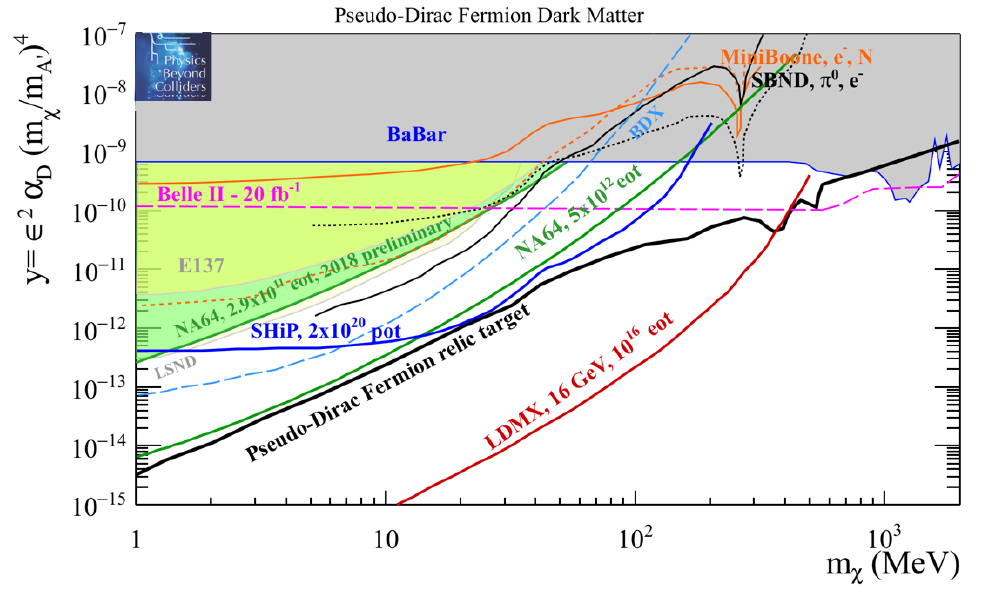}}
        \caption{\label{fig:PBC} Comparison of sensitivities of future collider and non-collider experiments for a model of light dark matter being produced via a new dark boson, showing the complementarity between different kinds of experiments in different ranges of dark matter particle mass and couplings with the dark boson. Taken from Ref.~\cite{Strategy:2019vxc,Beacham:2019nyx}.}
\end{figure}

The WIMP scenario shown in Figure~\ref{fig:DMCombinationDD} only represents a very simple benchmark in the landscape of theories on the nature of dark matter. Many other compelling explanations exist: for example the WIMP can be identified with the lightest, stable and invisible particle included in many supersymmetric models that also answer other outstanding questions of the Standard Model. Alternatives to the WIMP paradigm also exist, for example models where the dark matter particle is much lighter and has a mass below the GeV, see Fig.~\ref{fig:PBC} for a set of constraints of current and planned experiments. In these cases, searches at collider and direct detection experiments are complementary to searches at other planned dedicated accelerator experiments (e.g. beam dump experiments such as SHIP~\cite{Anelli:2015pba}, NA64~\cite{Banerjee:2016tad}, and LDMX~\cite{Akesson:2018vlm} among others), as well as underground experiments using novel sensor technologies (e.g. SENSEI \cite{Abramoff:2019dfb} and DAMIC~\cite{Aguilar-Arevalo:2019wdi}).

Axions (and axion-like particles)~\cite{PDGAxions, AxionEPNewsletter} also may be connected to solutions of the dark matter problem, being the DM particle candidate themselves or the mediators of the SM-DM interaction. Synergies between many different experiments and theoretical frameworks is evident in the case of those particles. Depending on the mass range and coupling of those particles, a discovery of such particles may occur  at the high-luminosity LHC, at lepton colliders (e.g. Belle II~\cite{Abe:2010gxa}) or at high-precision experiments that search for these particles directly (e.g. IAXO~\cite{Armengaud:2019uso} and ADMX~\cite{Braine:2019fqb}), or measure fundamental constants sensitive to fifth forces. Interaction between the members of these different experimental communities, as well as with theory and astrophysics, are needed to shape the future search program of these complementary experiments.

In general, connecting results and potential discoveries from different experiments within a coherent framework requires both particle and astroparticle physics theory involvement. This effort has been started by the LHC Dark Matter Working Group
\cite{DMWG}, where the Astroparticle community wishes to be further involved (see Ref.~\cite{APPECNews}). A parallel effort for non-WIMP, non-collider dark matter and dark sector searches is ongoing within the Physics Beyond Colliders Working Group.

A connection to nuclear physics in these and other (e.g. beam dump) dark matter experiments is needed to fully understand instrumental and beam backgrounds, as well as simulation. Particle and astroparticle experiments searching for dark matter also benefit from cross-talk in terms of instrumentation (e.g. sensors and cryogenics) and interpretation of results.

\section{\textit{Data firehoses} and shared solutions in high energy physics and astrophysics}

Another example of a common challenge for different fields is the ever-increasing volume of data available to different fields of research. Examples of current \textit{data firehoses} are the LHC, especially in light of the planned high-luminosity upgrade, and upcoming astrophysics surveys such as LSST~\cite{Ivezic:2008fe} and SKA~\cite{Bull:2018lat} to name but two. Similar challenges in data acquisition and recording are present in neutrino physics experiments, in the case of their supernova detection data streams. In all these cases, a fast and close-to-real-time analysis of the data is necessary, so that events of interest can be recorded or investigated further in a timely and cost-effective way, and common real-time analysis solutions are being investigated and deployed by multiple experiments.

Another point of contact is when software solutions are shared across fields, for example in the case of gravitational waves and high energy physics with the CernVM software appliance~\cite{CERNVMEPNewsletter} and the RUCIO distributed data management system that is in use by LHC experiments and will be adopted by neutrino experiments as well~\cite{RucioCERNNews}. 

\section{Collaborative efforts}

A number of platforms and fora exist at CERN and in Europe to facilitate cross-talk among different communities. In addition to the already-mentioned Dark Matter Working Group, there are (just to name a few) the recently inaugurated European Center for Astroparticle Physics currently hosted by CERN~\cite{EuCAPTCERNNews}, the European Science Cluster for Astronomy and Particle physics ESFRI research infrastructures project, the HEP Software Foundation to facilitate cooperation and common effort in software and computing, as well as the very successful CERN neutrino platform. 

\section{Conclusions}

The examples brought forward in this EPS-HEP contribution are only a very limited subset of how the particle, astroparticle and nuclear physics communities can be answering challenging scientific questions together. Other topics where synergies exist mentioned during the session were axion-like particles, the theory and experimental efforts bridging the gap between nuclear and high energy physics, and the opportunities offered by astrophysics experiments (e.g. Auger) spanning a much higher and complementary energy regime with respect to nuclear and particle physics experiments.

Since detector technologies are often common to different communities, the CERN expertise stemming from the current world-leading collider program can be reused. Moreover, data collection and analysis benefit from becoming faster, more efficient and more open: using versatile computing strategies and tools to solve diverse problems encourages common expertise that lasts beyond a single experiment.

In conclusion, there is the common wish that the European Strategy process will facilitate closer collaboration between the particle, astroparticle and nuclear physics communities, in a context where the design of detectors, data acquisition systems and computing are an integral part in our quest to understand the universe.

\end{document}